\documentclass[preprint,aps,pre,showpacs,preprintnumbers,amsmath,amssymb]
{revtex4}

\usepackage{graphicx}

\begin{document}

\title{Pattern formation in crystal growth under parabolic shear flow II}

\author{K. Ueno}

\email{ueno@riam.kyushu-u.ac.jp}

\affiliation{Research Institute for Applied Mechanics, Kyushu University, 6-1 Kasuga-koen, Kasuga, Fukuoka 816-8580, Japan}


\begin{abstract}
Wavy pattern of ice with a specific wavelength occurs during ice growth from a thin layer of undercooled water flowing down the surface of icicles or inclined plane. In the preceding paper [K. Ueno, Phys. Rev. E {\bf 68}, 021603 (2003)], we have found that restoring forces due to gravity and surface tension is a factor for stabilization of morphological instability of the solid-liquid interface. However, the mechanism for the morphological instability and stability of the solid-liquid interface has not been well understood. In the present paper, it is shown that a phase difference between fluctuation of the solid-liquid interface and distribution of heat flux at the deformed solid-liquid interface, which depends on the magnitude of the restoring forces, is a cause of the instability and stability of the interface. This mechanism is completely different from the usual Mullins-Sekerka instability due to diffusion and stabilization due to the Gibbs-Thomson effect.
\end{abstract}


\maketitle

\section{introduction}

\begin{figure}
\begin{center}
\includegraphics[width=8cm,height=8cm,keepaspectratio,clip]{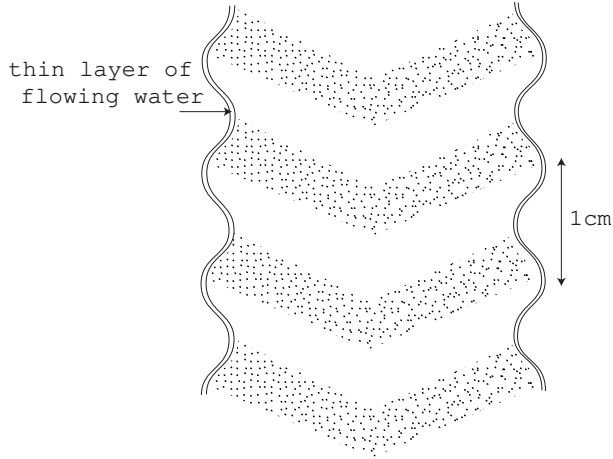}
\end{center}
\caption{Schematic diagram of vertical cross section of icicle covered with thin layer of flowing water and trapped many tiny air bubbles. Mean spacing between ribs is about 1 cm.}
\label{fig:sketchVp}
\end{figure}

\begin{figure}
\begin{center}
\includegraphics[width=8cm,height=8cm,keepaspectratio,clip]{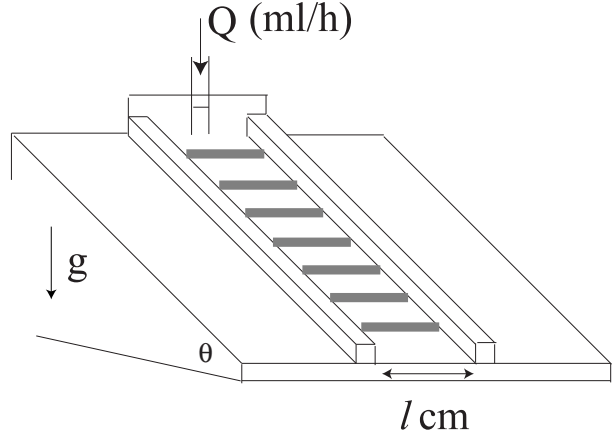}
\end{center}
\caption{Schematic diagram of ice growth from thin layer of undercooled water flowing down under the action of gravity on inclined plane with width $l$ cm and at angle $\theta$, which is set in a cold room below $0^{\circ}$ C. Water supply rate is $Q$ ml/h. Shaded regions with uniform spacing are nearly periodic wavy pattern of ice.}
\label{fig:inclinedplane}
\end{figure}

Ripple formation in sand induced by water shear flow \cite{1} and ribs and hollows formation on the surface of icicles covered with thin layer of flowing water (see Fig. \ref{fig:sketchVp} or Fig. 9A in Ref. \cite{2}) are well known phenomena in nature. The similar wavy pattern as ribs and hollows on icicles in nature can be experimentally produced during ice growth by continuously supplying a proper water $Q$ ml/h on an inclined plane with width $l$ and at angle $\theta$, set in cold room below $0^{\circ}$ C sketched in Fig. \ref{fig:inclinedplane}, and it is found that the mean wavelength of the wavy pattern of ice is given by $0.83/(\sin\theta)^{0.6\sim0.9}$ cm \cite{3}. In Fig. \ref{fig:inclinedplane}, the shaded regions with uniform spacing are protruded part of the wavy pattern of ice. Indeed, the spacing of the wavy pattern at $\theta=\pi/2$ is nearly equal to the mean spacing between ribs on the surface of icicles in nature. 

\begin{figure}
\begin{center}
\includegraphics[width=8cm,height=8cm,keepaspectratio,clip]{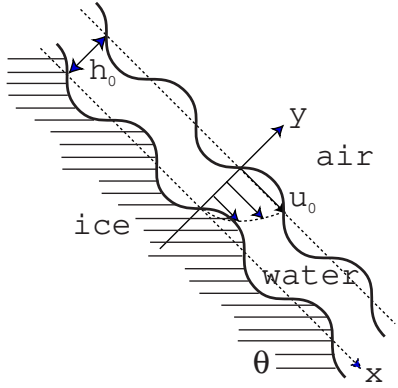}
\end{center}
\caption{Schematic diagram of vertical plane $(x,y)$ of inclined plane at angle $\theta$.}
\label{fig:diagram}
\end{figure}

In the previous works \cite{4,5}, a morphological instability of the solid-liquid interface during a crystal growth with mean velocity $\bar{V}$ from an undercooled thin liquid flowing down an inclined plane under the action of gravity as shown in Fig. \ref{fig:diagram} was investigated. We restricted ourselves to two dimensions in a vertical plane $(x,y)$, and for simplicity we assumed that the region of the crystal is semi-infinite. The $x$ axis is parallel to the inclined plane and the $y$ axis is normal to it. The parabolic shear flow \cite{10,11}
\begin{equation}
\bar{U}(y)=u_{0}\left\{2\frac{y}{h_{0}}-\left(\frac{y}{h_{0}}\right)^{2}\right\}\label{eq:intro1}
\end{equation}
is parallel to the $x$ axis and is bounded on one side by a liquid-air surface which is exposed by cold air below $0^{\circ}$ C. Here $u_{0}$ is the velocity at the free surface and $h_{0}$ is the mean thickness of the liquid, which can be expressed as \cite{10,11}:
\begin{equation}
h_{0}=\left(\frac{3\nu Q}{lg \sin\theta}\right)^{1/3},
\label{eq:intro2}
\end{equation}
where $g$ is the gravitational acceleration and $\nu$ is the kinematic viscosity. 

We discussed some differences between the dispersion relation for the fluctuation of the solid-liquid interface in our model and that in the Ogawa and Furukawa's model ( hereafter, we refer to thier model as $O$-$F$ model ). Our amplification rate $\sigma_{r}$ and phase velocity $v_{p}$ are given by \cite{4}
\begin{equation}
\sigma_{r}
=\frac{\bar{V}}{h_{0}}
\left[\frac{-\frac{3}{2}\alpha(\mu \rm Pe)
+\mu\left\{36-\frac{3}{2}\alpha(\mu \rm Pe)\right\}}{36+\alpha^{2}}
+n\mu
\frac{-\frac{7}{10}\alpha(\mu \rm Pe)-\alpha^{2}+\mu\left\{36-\frac{7}{10}\alpha(\mu \rm Pe)\right\}}{36+\alpha^{2}}\right],
\label{eq:intro3}
\end{equation}
\begin{equation}
v_{p}=-\frac{\bar{V}}{\mu}
\left[\frac{-\frac{1}{4}\alpha^{2}(\mu \rm Pe)
+\mu\left\{6\alpha+9(\mu \rm Pe)\right\}}{36+\alpha^{2}} 
+n\mu\frac{6\alpha-\frac{7}{60}\alpha^{2}(\mu \rm Pe)
+\mu\left\{6\alpha+\frac{21}{5}(\mu \rm Pe)\right\}}{36+\alpha^{2}}\right],
\label{eq:intro4}
\end{equation}
where $\mu=kh_{0}$, ${\rm Pe}=u_{0}h_{0}/\kappa_{l}$ is the Peclet number, $n=K_{s}/K_{l}$, $k$, $\kappa_{l}$, $K_{l}$ and $K_{s}$ being the wave number, the thermal diffusivity of the liquid, the thermal conductivity of the liquid and solid, respectively.
\begin{equation}
\alpha=2 \cot\theta h_{0}k+a^{2}h_{0}k^{3}
\label{eq:intro7}
\end{equation}
is restoring forces due to gravity and surface tension acting on the liquid-air surface, $a=\sqrt{2\gamma/(\rho_{l}g \sin\theta)}$ being the capillary constant associated with the surface tension $\gamma$ of the liquid-air surface. While, by using our notations, the $O$-$F$ model gives \cite{5} 
\begin{eqnarray}
\sigma_{r} = \bar{V}k\frac{1-\frac{239}{10080}(\mu {\rm Pe})^2}{\left\{1-\frac{239}{10080}(\mu {\rm Pe})^{2}\right\}^{2}+\left\{\frac{5}{12}\mu {\rm Pe}\right\}^{2}},
\label{eq:intro5}
\end{eqnarray}
\begin{eqnarray}
v_{p} 
=\bar{V}\frac{\frac{5}{12}\mu {\rm Pe}}{\left\{1-\frac{239}{10080}(\mu {\rm Pe})^{2}\right\}^{2}+\left\{\frac{5}{12}\mu {\rm Pe}\right\}^{2}}.
\label{eq:intro6}
\end{eqnarray}

A critical differece of these dispersion relations is that our result includes the restoring forces $\alpha$, while the result of $O$-$F$ model does not include it. The difference of $\sigma_{r}$ results in different dependence of wavelength $\lambda_{\rm max}$, at which $\sigma_{r}$ takes a maximum value, on $\theta$. Our result is better agreement with $\lambda_{\rm mean}$ obtained by experiment \cite{3} than the result of $O$-$F$ model (see Fig. 4 in Ref. \cite{4}). The difference of $v_{p}$ results in different direction of migration of the solid-liquid interface. Our results predict that it moves upward with about $0.6\bar{V}$. Indeed, our prediction is consistent with the observation that many tiny air bubbles trapped in just upstream region of any protruded part migrate in the upward direction during growth as shown in Fig. \ref{fig:sketchVp}. On the other hand, the result of $O$-$F$ model predicts that it moves downward with about $0.5\bar{V}$. Since our calculations based on a linear stability analysis and the $O$-$F$'s calculations are different, the cause leading to these different results was not clarified in the previous paper. In addition to the presence of $\alpha$ or not, it was only suggested that these differences may be due to the difference of boundary conditions of the temperature at the solid-liquid interface and that of the liquid-air surface. In Sec. II, we confirm it by deriving $O$-$F$ model from our formulation.

According to the result of $O$-$F$ model, the instability of the solid-liquid interface occures by the Laplace instability due to the thermal diffusion into the air, and its instability is suppressed by the effect of fluid flow, which makes the temperature distribution in the thin water layer uniform \cite{5}. However, this qualitative interpretation does not enable to explain the migration of the interface, and the role of fluid flow on the temperature field in the liquid is not quantitatively clear. In our previous paper, although it was suggested that the restoring forces are indeed an important factor for stabilization of morphological instability of the solid-liquid interface (see Fig. 2 in Ref. \cite{4}), the mechanism was not well understood. In Sec. III, therefore, we clarify the morphological instability and stability mechanism of the solid-liquid interface, and we present a physical mechanism for migration of the interface in the upstream direction. Conclusion is given in Sec. IV. 

\section{Difference in boundary conditions}
The general solution for the amplitude of perturbed temperature in the liquid is\begin{equation}
g_{l}(z)=B_{1}\phi_{1}(z)+B_{2}\phi_{2}(z)+i\mu {\rm Pe} \int_{0}^{z}\left\{\phi_{2}(z)\phi_{1}(z')-\phi_{1}(z)\phi_{2}(z')\right\}
f(z')dz'\bar{G}_{l}\zeta_{k},
\label{eq:pre1}
\end{equation}
where $\phi_{1}(z)$ and $\phi_{2}(z)$ are given by Eqs. (81) and (82) in Ref. \cite{4}, $\bar{G}_{l}$ is the unperturbed temperature gradient in the liquid, and $z=1-y/h_{0}$.

The first difference of our model and $O$-$F$ model is the form of the amplitude $f$ of perturbed stream function in Eq. (\ref{eq:pre1}). The solution of $f$ was determined under the same hydrodynamic boundary conditions in the two models, but the difference of the order estimate of the restoring force $\alpha$ led to different solutions. Our result in the long wavelength approximation $\mu \ll 1$ gives 
\begin{eqnarray}
f(z)
&=&\frac{1}{6-i\alpha}(-6+i\alpha z+6z^{2}-i\alpha z^{3}) \nonumber \\
& &-\frac{\mu {\rm Re}\alpha}{210(6-i\alpha)^{2}}
\left\{144+(-174+5i\alpha)z-144z^{2} \right.
\nonumber \\
& &\left.+(210-11i\alpha)z^{3}+(-42+7i\alpha)z^{5}+(6-i\alpha)z^{7}\right\},
\label{eq:pre2}
\end{eqnarray}
where ${\rm Re}=u_{0}h_{0}/\nu$ is the Reynolds number.
While the result of $O$-$F$ model's gives
\begin{equation}
f(z)=z^2-1.
\label{eq:pre3}
\end{equation}
From the kinematic condition at the liquid-air surface, the reletion between the amplitude $\zeta_{k}$ of the perturbed solid-liquid interface and the amplitude $\xi{_k}$ of the perturbed liquid-air surface was given by $\xi_{k}=-f|_{z=0}\zeta_{k}$ \cite{4}. Equation (\ref{eq:pre2}) includes $\alpha$ which makes the amplitude and phase of the liquid-air surface change, it deforms depending on the wavelength of the fluctuation of solid-liquid interface. While, Eq. (\ref{eq:pre3}) does not include it. Then, the liquid-air surface fluctuates with the same amplitude as the solid-liquid interface and phase shift of each interface does not occur. 

The second difference between our model and $O$-$F$ model is originated from the difference of $B_{1}$ and $B_{2}$ in Eq. (\ref{eq:pre1}),  which depend on the choice of the boundary conditions of the temperature at the solid-liquid interface or the liquid-air surface. The two differences lead to a critical difference of the dispersion relations for the fluctuation of the solid-liquid interface and of the corresponding mechanism of instability and stability of it. The difference of the mechanism will be discussed in detail in the next section. Here we describe the essential difference of the thermodynamic boundary conditions in the two models, and we correct the solution Eq. (86) of unperturbed temperature of the air obtained in Ref. \cite{4}. 


The first essential difference is the continuity of the temperature at the perturbed solid-liquid interface $\zeta(t,x)=\zeta_{k}\exp[\sigma t+ikx]$, where $\sigma=\sigma_{r}+i\sigma_{i}$ and $t$ is time. In our model,
\begin{equation}
T_{l}|_{y=\zeta}=T_{s}|_{y=\zeta}=T_{m}+\Delta T,
\label{eq:Tb1}
\end{equation}
where $T_{m}$ is the equilibrium melting temperature. Here we assume that a deviation from the equilibrium melting temperature, $\Delta T$, is of order $\zeta_{k}$ and this corresponds to $G(k)\zeta$ in the previous paper \cite{4}. If we regard  $\Delta T$ as the melting temperature depression due to the Gibbs-Thomson effect \cite{20}, we can neglect it as far as we are concerned with the wavelength of the wavy pattern observed on the surface of icicles or the inclined plane. Then, the temperature at the solid-liquid interface in a pure substance must be the equilibrium melting temperature. Therefore, in the $O$-$F$ model,  
\begin{equation}
T_{l}|_{y=\zeta}=T_{s}|_{y=\zeta}=T_{m}.
\label{eq:Tb2}
\end{equation}
However, we suggested that there can be a deviation from the equilibrium melting temperature, which can not be determined a priori and is determined after we determine the solution for the perturbed temperature in the liquid \cite{4}. 

The second essential difference is the continuity of the temperature at the liquid-air surface $\xi(t,x)=h_{0}+\xi_{k}\exp[\sigma t+ikx]$. In our model,
\begin{equation}
T_{l}|_{y=\xi}=T_{a}|_{y=\xi}=T_{la},
\label{eq:Tb3}
\end{equation}
where $T_{la}$ is a temperature at the liquid-air surface. While, in the $O$-$F$model, 
\begin{equation}
T_{l}|_{y=\xi}=T_{a}|_{y=\xi}.
\label{eq:Tb4}
\end{equation}
Equation (\ref{eq:Tb3}) indicates that the temperature of the liquid-air surface remains a constant $T_{la}$ after deformation of the liquid-air surface. While, Eq. (\ref{eq:Tb4}) shows that the temperature at the deformed liquid-air surface is not necessary to remain a constant value. 

From the heat conservations Eq. (94) at the unperturbed liquid-air surface and Eq. (101) at the unperturbed solid-liquid interface in Ref. \cite{4}, we obtain $T_{la}=T_{\infty}+L/C_{pa}$, $T_{\infty}$, $L$ and $C_{pa}$ being the ambient air temperature, the latent heat per unit volume and the specific heat of the air at constant pressure, respectively. In this case, nothing determines the growth velocity $\bar{V}$, which may take any value. This problem also occur in the absence of flow \cite{20}. There is a more serious problem. The value of $L/C_{pa}$ is about $2.54 \times 10^{3}$ K. Then, for the real values of $T_{\infty}$ observed \cite{2,3}, the value of $T_{la}$ becomes larger than the melting temperature. This problem is originated from the unperturbed solution (86) of the temperature $\bar{T}_{a}$ of the air in Ref. \cite{4}. Here we briefly improve the unperturbed solution. 

Under the boundary conditions $\bar{T}_{a}=T_{la}$ at $y=h_{0}$ and $\bar{T}_{a}=T_{\infty}$ at $y=h_{0}+l_{a}$, we assume that by neglecting the term $\bar{V}$ in Eq. (85) in Ref. \cite{4} the approximate solution of the unperturbed part near the liquid-air surface is given by
\begin{equation}
\bar{T}_{a}(y)=T_{la}-\bar{G}_{a}(y-h_{0}),
\label{eq:Tb5}
\end{equation}
where $\bar{G}_{a}=(T_{la}-T_{\infty})/l_{a}$ is the unperturbed temperature gradient in the air at $y=h_{0}$, $l_{a}$ being a length of thermal diffusion layer ahead of the liquid-air surface. We note that this $l_{a}$ is not $l_{a}=\kappa_{a}/\bar{V}$ in the previous paper \cite{4}. From the heat conservation $K_{l}\bar{G}_{l}=K_{a}\bar{G}_{a}$ at the unperturbed liquid-air surface, where $K_{a}$ is the thermal conductivity of the air,  
$T_{la}$ is obtained as
\begin{equation}
T_{la}=\frac{T_{m}+\frac{K_{a}}{K_{l}}\frac{h_{0}}{l_{a}}T_{\infty}}
{1+\frac{K_{a}}{K_{l}}\frac{h_{0}}{l_{a}}}.
\label{eq:Tb6}
\end{equation}
If the values of $T_{\infty}$, $h_{0}$ and $l_{a}$ are given, the value of $T_{la}$ in Eq. (\ref{eq:Tb3}) is determined. 
Substituting Eq. (\ref{eq:Tb6}) into the heat conservation $L\bar{V}=K_{l}\bar{G}_{l}$ at the unperturbed solid-liquid interface yields
\begin{equation}
\bar{V}=\frac{K_{l}}{Lh_{0}}\frac{T_{m}-T_{\infty}}{1+\frac{K_{l}}{K_{a}}\frac{l_{a}}{h_{0}}}.
\label{eq:Tb7}
\end{equation}
Since $K_{l}/K_{a} \gg 1$, Eq. (\ref{eq:Tb7}) can be approximated as
\begin{equation}
\bar{V} \approx \frac{K_{a}}{L}\frac{T_{m}-T_{\infty}}{l_{a}}.
\label{eq:Tb8}
\end{equation}
We note that $\bar{V}$ does not depend on $h_{0}$ which varies with $Q$. Therefore, $\bar{V}$ is not affected by change of $Q$. This agrees with observations that as increasing water supply rate growth velocity of diameter of icicles is almost constant \cite{2}. In the previous paper, we determined the value of $T_{la}$ from only Eq. (115) in Ref. \cite{4} using the actual observed value of $\bar{V}$. However, the expression (115) is not appropriate for determinig $\bar{V}$ because it changes with $h_{0}$.


Under the boundary conditions (\ref{eq:Tb1}) and (\ref{eq:Tb3}), we obtain \cite{4}
\begin{equation}
B_{1}=-f|_{z=0}\bar{G}_{l}\zeta_{k},
\hspace{1cm}
B_{2}=\mu B_{1},
\label{eq:dis1}
\end{equation}
where $f|_{z=0}$ is given by Eq. (125) in Ref. \cite{4}.
Then, the dispersion relation for the fluctuation of the solid-liquid interface becomes 
\begin{equation}
\sigma = \frac{\bar{V}}{h_{0}}\left\{\frac{dH_{l}}{dz}\Big|_{z=1}
          +n\mu\left(H_{l}|_{z=1}-1\right)\right\},
\label{eq:dis2}
\end{equation}
where 
\begin{equation}
H_{l}(z)=-f|_{z=0}\left\{\phi_{1}(z)+\mu\phi_{2}(z)\right\}
+i\mu {\rm Pe}\int_{0}^{z}\left\{\phi_{2}(z)\phi_{1}(z')-\phi_{1}(z)\phi_{2}(z')\right\}
f(z')dz'.
\label{eq:dis3}
\end{equation}
The real and imaginary part of Eq. (\ref{eq:dis2}) give Eqs. (\ref{eq:intro3}) and (\ref{eq:intro4}) by approximating Eqs. (97), (119) and (120), respectively \cite{4}. In the previous paper, nevertheless we had inappropriate base state of the temperature of the air, we obtained reasonable results compatible with experiments and observations. This is because the change of $\bar{V}$ in Eq. (\ref{eq:Tb8}) by $T_{\infty}$ affects the magnitude of $\sigma_{r}$ but does not make change the characteristic wavelength of the wavy pattern determind from the maximum point of $\sigma_{r}$. 

On the other hand, under the boundary conditions (\ref{eq:Tb2}) and (\ref{eq:Tb4}), we obtain a different
\begin{equation}
B_{1}=\frac{1-i\mu {\rm Pe}I|_{z=1}+\mu \left(1-\frac{K_{a}}{K_{l}}\right)f|_{z=0}\phi_{2}|_{z=1}}{\phi_{1}|_{z=1}+\mu\frac{K_{a}}{K_{l}}\phi_{2}|_{z=1}}\bar{G}_{l}\zeta_{k},
\label{eq:dis4}
\end{equation}
\begin{equation}
B_{2}=\frac{-\mu \left(1-\frac{K_{a}}{K_{l}}\right)f|_{z=0}\phi_{1}|_{z=1}+\mu\frac{K_{a}}{K_{l}}\left(1-i\mu {\rm Pe}I|_{z=1}\right)}{\phi_{1}|_{z=1}+\mu\frac{K_{a}}{K_{l}}\phi_{2}|_{z=1}}\bar{G}_{l}\zeta_{k},
\label{eq:dis5}
\end{equation}
where
\begin{equation}
I(z)\equiv
\int_{0}^{z}\left\{\phi_{2}(z)\phi_{1}(z')-\phi_{1}(z)\phi_{2}(z')\right\}f(z')dz'. 
\label{eq:dis6}
\end{equation}
Then, the dispersion relation becomes 
\begin{equation}
\sigma = \frac{\bar{V}}{h_{0}}\frac{dH_{l}}{dz}\Big|_{z=1},
\label{eq:dis7}
\end{equation}
where
\begin{equation}
H_{l}(z)=\frac{1}{\phi_{1}|_{z=1}}\left[\left\{1-i\mu {\rm Pe}I|_{z=1}\right\}\phi_{1}(z)+\mu f|_{z=0}\left\{\phi_{2}|_{z=1}\phi_{1}(z)-\phi_{1}|_{z=1}\phi_{2}(z)\right\}\right]+i\mu {\rm Pe}I(z).
\label{eq:dis8}
\end{equation}
In Eq. (\ref{eq:dis8}), we have omitted the term $\mu K_{a}/K_{l} \ll 1$.
In particular, when puttinig $\alpha=0$ in $f$ in Eq. (\ref{eq:dis8}), we recover Eqs. (\ref{eq:intro5}) and (\ref{eq:intro6}) from the real and imaginary part of Eq. (\ref{eq:dis7}), respectively. 

\begin{figure}
\begin{center}
\includegraphics[width=8cm,height=8cm,keepaspectratio,clip]{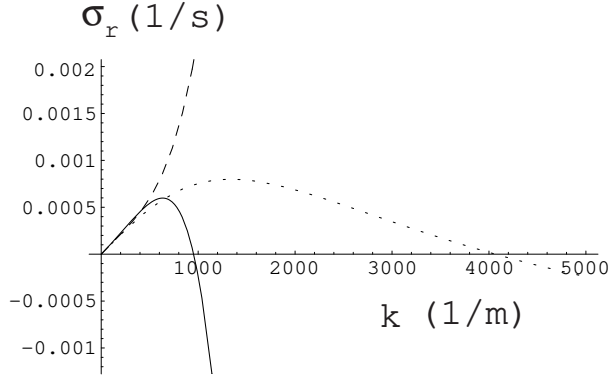}
\end{center}
\caption{The amplification rate $\sigma_{r}$ vs the wave number $k$ for $\bar{V}=10^{-6}$ m/s, $Q=160$ ml/h, and $\theta=\pi/2$. Solid line: Re[Eq. (\ref{eq:dis2})]. Dashed line: Re[Eq. (\ref{eq:dis7})]. Dotted line: Re[Eq. (\ref{eq:dis7})] ($\alpha=0$).}
\label{fig:growthrate}
\end{figure}

\begin{figure}
\begin{center}
\includegraphics[width=8cm,height=8cm,keepaspectratio,clip]{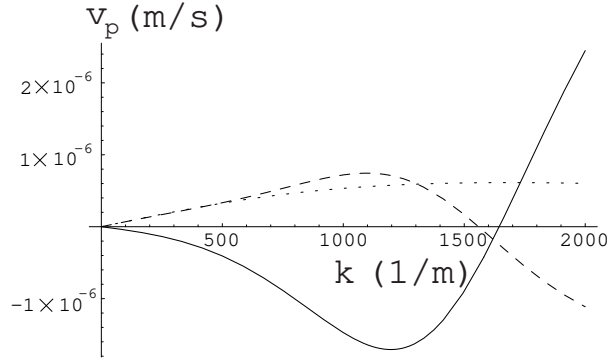}
\end{center}
\caption{Phase velocity $v_{p}=-\sigma_{i}/k$ vs the wave number $k$ for $\bar{V}=10^{-6}$ m/s, $Q=160$ ml/h, and $\theta=\pi/2$. Solid line: -Im[Eq. (\ref{eq:dis2})]/k. Dashed line: -Im[Eq. (\ref{eq:dis7})]/k. Dotted line: -Im[Eq. (\ref{eq:dis7})]/k ($\alpha=0$).}
\label{fig:phasevelocity}
\end{figure}


In the absence of flow, if we regard $\Delta T$ in Eq. (\ref{eq:Tb1}) as the Gibbs-Thomson effect, and by replacing the amplitude relation $\xi_{k}=-f|_{z=0}\zeta_{k}$ \cite{4} with
\begin{equation}
\xi_{k}=\exp(-\mu)\left(1-d_{0}\frac{\kappa_{l}}{\bar{V}}k^{2}\right)\zeta_{k},
\label{eq:MS1}
\end{equation}
we obtain
\begin{equation}
H_{l}(z)=\exp[-\mu(1-z)]\left(1-d_{0}\frac{\kappa_{l}}{\bar{V}}k^{2}\right).
\label{eq:MS2}
\end{equation}
Then, Eq. (\ref{eq:dis2}) obtained from the boundary conditions (\ref{eq:Tb1}) and (\ref{eq:Tb3}) reduces to the dispersion relation in the Mullins-Sekerka theory \cite{20}:
\begin{equation}
\sigma_{r}=\bar{V}k\left[1-d_{0}\frac{\kappa_{l}}{\bar{V}}(1+n)k^{2}\right],
\label{eq:MS3}
\end{equation}
and $v_{p}=0$,
where $d_{0}=T_{m}\Gamma C_{p}/L^{2}$ is the capillary length, $\Gamma$ being the solid-liquid interface tension \cite{20}.
We note that we can not recover the dispersion relation in the Mullins-Sekerka theory from the boundary conditions (\ref{eq:Tb2}) and (\ref{eq:Tb4}) even if we add the Gibbs-Thomson effect to Eq. (\ref{eq:Tb2}). 

The solid lines, dashed lines and dotted lines ($\alpha=0$) in Figs. \ref{fig:growthrate} and \ref{fig:phasevelocity} represent $\sigma_{r}$ and $v_{p}=-\sigma_{i}/k$ obtained from Eqs. (\ref{eq:dis2}) and (\ref{eq:dis7}) with the use of Eqs. (97), (119) and (120), respectively. Here Re and Im denote the real and imaginary part of arguments, respectively, and note that Re is not the Reynolds number in the following discussion. It is found from the dashed line in Fig. \ref{fig:growthrate} that $\sigma_{r}$ is always positive, therefore, it is not impossible to get the characteristic wavelength observed on the surface of icicles or the inclined plane. The dotted line in Fig. \ref{fig:growthrate} has a maximum point of $\sigma_{r}$ at a specific wave number. However, since this result of $O$-$F$ model does not consider the effect of restoring forces due to gravity and surface tension on the liquid-air surface, deviation from the experiment is large as shown in the closed triangles in Fig. 4 in Ref. \cite{4}. As shown in the solid line in Fig. \ref{fig:phasevelocity}, the direction of phase velocity in our model is negative in our interest wave number region. On the other hand, as shown in the dashed and dotted lines in Fig. \ref{fig:phasevelocity}, the direction of phase velocity in the models different from ours is positive in our interest wave number region. There is no evidence to support this prediction.

From these considerations, in order to explain the experiments and observations, and in the absence of flow, to recover the dispersion relation in the Mullins-Sekerka theory, the boundary conditions (\ref{eq:Tb1}) and (\ref{eq:Tb3}) seem to be most appropriate.


The deviation from the equilibrium melting temperature in our model is given by \cite{4}
\begin{equation}
\Delta T=(H_{l}|_{z=1}-1)\bar{G}_{l}\zeta,
\label{eq:shift1}
\end{equation}
where $H_{l}$ is given by Eq. (\ref{eq:dis3}). In the absence of flow, instead of Eq. (\ref{eq:dis3}), if we apply Eq. (\ref{eq:MS2}) to Eq. (\ref{eq:shift1}), we recover the Gibbs-Thomson effect, $\Delta T=-d_{0}l_{l}k^{2}\bar{G}_{l}\zeta$, where $l_{l}=\kappa_{l}/\bar{V}$. We note that as $k \rightarrow 0$, the solid-liquid interface is flat and both $H_{l}|_{z=1}$ in Eqs. (\ref{eq:dis3}) and (\ref{eq:MS2}) approach 1, therefore, $\Delta T$ vanishes. This indicates that the deviation from the equilibrium melting temperature in our systems is not induced by only the effect of shear flow Eq. (\ref{eq:intro1}).

We note the difference of the characteristic length scale in our problem with flow and in the Mullins-Sekerka theory. The capillary length $d_{0}$ associated with the solid-liquid interface tension is a microscopic length of order angstroms, while $l_{l}$ is usually macroscopic. Therefore, the Gibbs-Thomson effect acts effectively on the micrometer scale \cite{20}. On the other hand, the restoring forces in Eq. (\ref{eq:intro7}) include the capillary constant $a$ assosiated with the surface tension of the liquid-air surface, which is 3.9 mm for water at $\theta=\pi/2$, and the thickness $h_{0}$ of the liquid, which is about $10^{-4}$ m. Then, the effect of restoring forces is more effective for longer wavelength fluctuation compared to the length scale the Gibbs-Thomson effect is effective. We can neglect $\Delta T$ due to the Gibbs-Thomson effect but not neglect it due to the restoring forces. Indeed, the effect of $\Delta T$ is reflected in the second term of Eq. (\ref{eq:dis2}). 

Local equilibrium thermodynamics assumes that the equations of state retain the same form out of equilibrium as in equilibrium, but with a local meaning \cite{21}. Then, the flow does not change the equations of state and the equilibrium melting temperature is determined from equality of the chemical potentials of the crystal and liquid. This local equilibrium hypothesis is implicitly used in the $O$-$F$ model and leads to Eq. (\ref{eq:Tb2}). However, if we impose the boundary condition (\ref{eq:Tb2}) in our systems, we can not get desired results compatible with experiments and observations. The restoring forces $\alpha$ have a purely hydrodynamic origin, but there is an interplay of both thermodynamic and hydrodynamic effect in $\Delta T$ determined from Eq. (\ref{eq:dis3}), which is reflected in the term $\alpha$Pe in $H_{l}$. 

The thermodynamics of fluids under shear flow is an active and very challenging topic in modern nonequilibrium thermodynamics and statistical mechanics \cite{21}. This is a field with many open questions. A decisive step in the thermodynamic understanding of $\Delta T$ is to formulate a free energy or a chemical potential depending explicitly on the characteristics of the flow. However, it is reported that the coexistence of the crystal and liquid under shear flow cannot be accounted for by invoking a nonequilibrium analogue of the chemical potential \cite{22}. We have seen that if we impose the equilibrium melting temperature at the solid-liquid interface, we can not get consistent results with actual experiments and observations. Therefore, the approach by the local equilibrium hypothesis may also be insufficient to deal with our systems. Whether the coexistence of the crystal and liquid under shear flow in the present case can be accounted for by introducing a nonequilibrium analogue of the chemical potential is left for future studies. 

\section{Mechanism of instability and stability of the solid-liquid interface}

In the preceding section, we have studied some models which differ in the form of boundary conditions. In this section, we present in detail the mechanism of instability and stability of the solid-liquid interface for our model. For other models, we give a brief discussion.

As shown in Fig. \ref{fig:hfluxk634,hfluxk953,hfluxk1200}, we consider a small perturbation of the solid-liquid interface with $\zeta_{k}/h_{0}=0.1$ at time $t$: 
\begin{equation}
{\rm Im}\left[\frac{\zeta}{h_{0}}\right]
=\sin [k(x-v_{p}t)]\frac{\zeta_{k}}{h_{0}}\exp(\sigma_{r}t).
\label{eq:mech1}
\end{equation}
Using $\xi_{k}=-f|_{z=0}\zeta_{k}$ \cite{4}, the corresponding perturbation of the liquid-air surface is
\begin{equation}
{\rm Im}\left[\frac{\xi}{h_{0}}\right]
={\rm Im}\left[-f|_{z=0}\exp[ik(x-v_{p}t)]\right]\frac{\zeta_{k}}{h_{0}}\exp(\sigma_{r}t).
\label{eq:mech2}
\end{equation}

Next we define the perturbation of heat flux into the liquid $q_{l}$ and solid $q_{s}$ at the perturbed solid-liquid interface and the perturbation of heat flux into the air $q_{a}$ at the perturbed liquid-air surface as follows, respectively,
\begin{eqnarray}
q_{l} &\equiv& {\rm Im}\left[-K_{l}\frac{\partial T'_{l}}{\partial y}\Big|_{y=\zeta}\right] \nonumber \\
&=& {\rm Im}\left[\frac{dH_{l}}{dz}\Big|_{z=1}\exp[ik(x-v_{p}t)]\right]K_{l}\bar{G}_{l}\frac{\zeta_{k}}{h_{0}}\exp(\sigma_{r}t),
\label{eq:mech3}
\end{eqnarray}
\begin{eqnarray}
q_{s} &\equiv & {\rm Im}\left[-K_{s}\frac{\partial T'_{s}}{\partial y}\Big|_{y=\zeta}\right] \nonumber \\
&=& -{\rm Im}\left[(H_{l}|_{z=1}-1)\exp[ik(x-v_{p}t)]\right]n\mu K_{l}\bar{G}_{l}\frac{\zeta_{k}}{h_{0}}\exp(\sigma_{r}t),
\label{eq:mech4}
\end{eqnarray}
\begin{eqnarray}
q_{a} &\equiv& {\rm Im}\left[-K_{a}\frac{\partial T'_{a}}{\partial y}\Big|_{y=\xi}\right] \nonumber \\
&=& {\rm Im}\left[-f|_{z=0}\exp[ik(x-v_{p}t)]\right]\mu K_{l}\bar{G}_{l}\frac{\zeta_{k}}{h_{0}}\exp(\sigma_{r}t),
\label{eq:mech5}
\end{eqnarray}
where we use Eq. (\ref{eq:dis3}) for $H_{l}(z)$.
We note the direction of heat flow. If $q_{l}>0$ or $q_{s}<0$, the latent heat is released away from the solid-liquid interface into each phase. Conveniently, the distribution of $q_{l}-q_{s}$ represented by the bottom dashed line and $q_{a}$ represented by the up dashed line are superimposed on Fig. \ref{fig:hfluxk634,hfluxk953,hfluxk1200} with magnification of 0.1 to see phase difference between the fluctuation of the solid-liquid interface, liquid-air surface and distribution of heat flux at the respective interfaces.

\begin{figure}
\begin{center}
\includegraphics[width=8cm,height=8cm,keepaspectratio,clip]{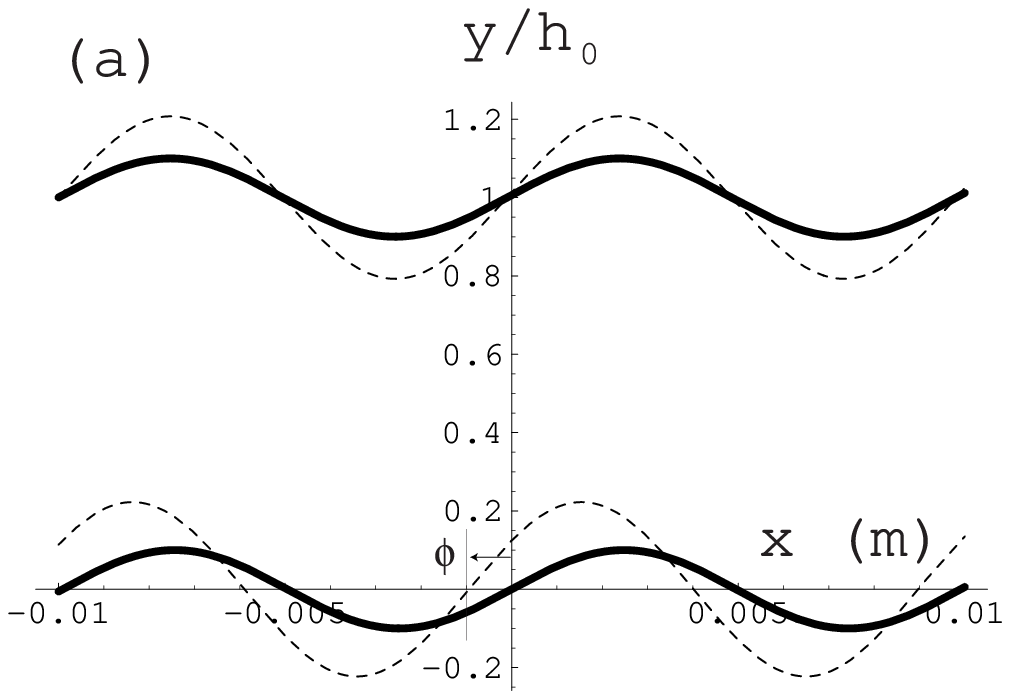}
\\[5mm]
\includegraphics[width=8cm,height=8cm,keepaspectratio,clip]{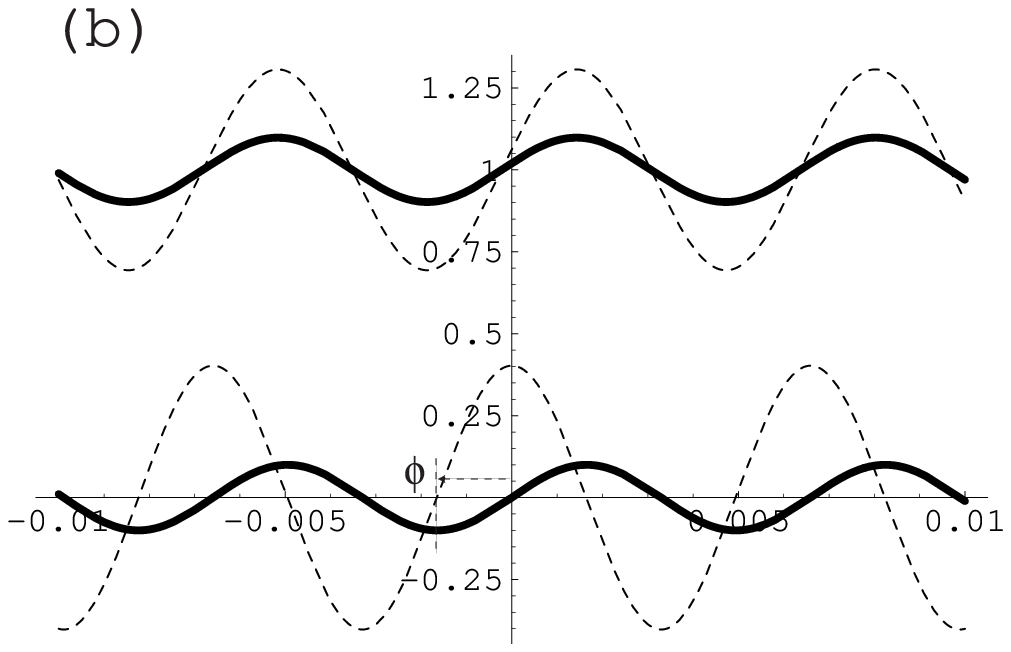}
\\[5mm]
\includegraphics[width=8cm,height=8cm,keepaspectratio,clip]{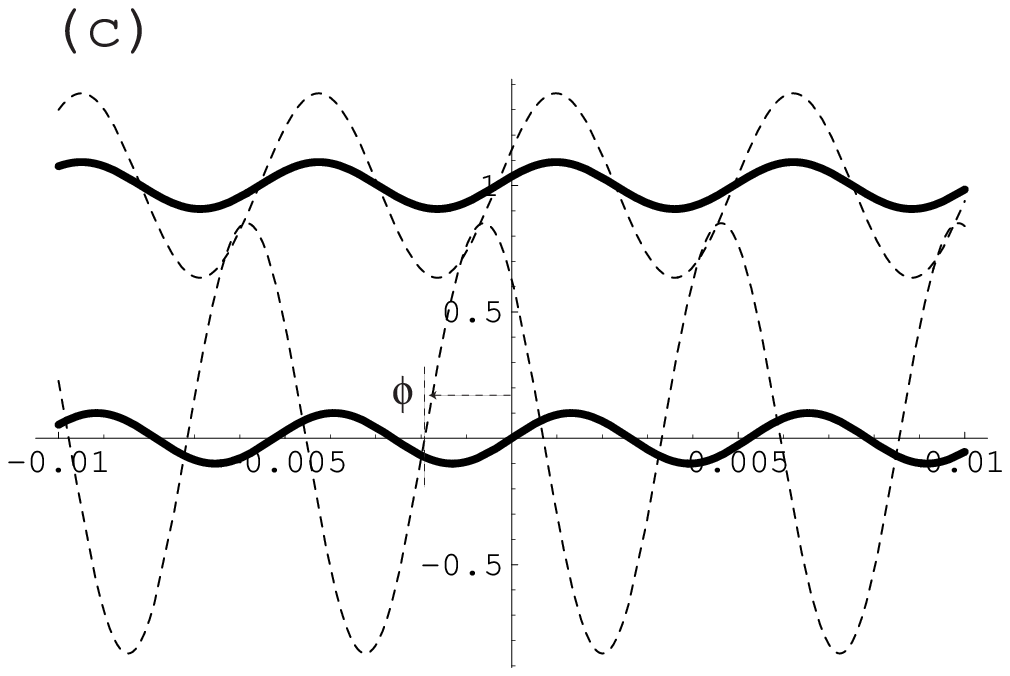}
\end{center}
\caption{Schematic illustration of the fluctuation of the solid-liquid interface ${\rm Im}[\zeta/h_{0}]$ (bottom thick solid lines), the liquid-air surface $1+{\rm Im}[\xi/h_{0}]$ (up thick solid lines), heat flux $q_{l}-q_{s}$ (bottom dashed lines) at ${\rm Im}[\zeta/h_{0}]$, and heat flux $q_{a}$ (up dashed lines) at $1+{\rm Im}[\xi/h_{0}]$ for (a) $k=634/{\rm m}$, (b) $k=953/{\rm m}$, and (c) $k=1200/{\rm m}$. $\phi$ is phase shift of heat flux $q_{l}-q_{s}$ against the solid-liquid interface.}
\label{fig:hfluxk634,hfluxk953,hfluxk1200}
\end{figure}

Figure. \ref{fig:hfluxk634,hfluxk953,hfluxk1200} (a) shows the configurations at a wave number in the unstable region $\sigma_{r}>0$ of the solid line in Fig. \ref{fig:growthrate}. The heat flux $q_{a}$ at the liquid-air surface is large at any protruded part of surface pointing into the air, at which the temperature gradient increases so that heat transfer by thermal diffusion into the air is more effective. Since the value of $\alpha$ is small in such low wave number region, that is, the effect of restoring forces on the liquid-air surface is small, the liquid-air surface fluctuates with almost the same amplitude as the solid-liquid interface, and the phase difference between ${\rm Im}[\xi/h_{0}]$ and ${\rm Im}[\zeta/h_{0}]$ is negligible. Therefore, this seems to result in faster cooling and hence freezing to promote at the protruded part of the solid-liquid interface. This picture of destabilization appears to be the same as the Mullins-Sekerka instability \cite{20}. However, we note that the maximum point of heat flux $q_{l}-q_{s}$ is shifted to the upstream direction by $\phi$ against the solid-liquid interface. This indicates that the interface grows faster in just the upstream region of any protruded part, in which  $q_{l}-q_{s}$ is large compared to the mean heat flux $K_{l}\bar{G}_{l}$. On the other hand, in the downstream region of any protruded part, the interface tends to melt back because $q_{l}-q_{s}$ is small compared to $K_{l}\bar{G}_{l}$. Therefore, the solid-liquid interface not only grows unstably but also moves in the upstream direction, which is consistent with the direction of phase velocity of the interface predicted by Eq. (\ref{eq:intro4}). It also support the observation that many tiny air bubbles dissolved in the thin flowing water are trapped in just upstream region of any protruded part on a growing icicle and its region migrates in the upward direction during growth as shown in the dotted regions in Fig. \ref{fig:sketchVp}. We can not explain this observation by usual Mullins-Sekerka instability \cite{20} or Laplace instability \cite{5} due to diffusion. 

Figure \ref{fig:hfluxk634,hfluxk953,hfluxk1200} (b) shows the configurations at the wave number at the neutral stability point $\sigma_{r}=0$ of the solid line in Fig. \ref{fig:growthrate}. Shift of the maximum point of heat flux $q_{l}-q_{s}$ is larger than that in Fig. \ref{fig:hfluxk634,hfluxk953,hfluxk1200} (a). Figure \ref{fig:hfluxk634,hfluxk953,hfluxk1200} (c) shows the configurations at a wave number in the stable region $\sigma_{r}<0$ of the solid line in Fig. \ref{fig:growthrate}. Since the value of $\alpha$ increases as increasing the wave number, the surface ${\rm Im}[\xi/h_{0}]$ and heat flux $q_{a}$ is slightly shifted to the upstream direction, and the phase difference between ${\rm Im}[\zeta/h_{0}]$ and $q_{l}-q_{s}$ becomes larger than that in Fig. \ref{fig:hfluxk634,hfluxk953,hfluxk1200} (b). Any protruded part of the perturbed solid-liquid interface melts back because $q_{l}-q_{s}$ is small, and any depression part of the interface grows because $q_{l}-q_{s}$ is large. Therefore, the flatness of the solid-liquid interface is restored, that is, the solid-liquid interface is stabilized by large phase shift of distribution of heat flux by large restoring forces. This stabilizing mechanism on such long length scales shown in Fig. \ref{fig:hfluxk634,hfluxk953,hfluxk1200} is different from the Gibbs-Thomson effect. 

If we apply the picture of the Mullins-Sekerka instability to the bottom thick solid lines in Fig. \ref{fig:hfluxk634,hfluxk953,hfluxk1200}, the deformed isotherms get closer to each other ahead of the bump of the solid-liquid interface. The temperature gradient, and therefore the heat flux, increases, which increases the rate of production of latent heat. Therefore, the bump must get amplified \cite{20}. However, we note that this picuture is true only in the absence of flow. In the Mullins-Sekerka instability, it is diffusion which destabilizes the planar front. In the presence of flow, the perturbed temperature field in the liquid is affected by the flow field, which varies depending on the magnitude of the restoring forces acting on the liquid-air surface. We can not determine a priori where the temperature gradient or heat flux is large until we solve the equation of the temperature field for a given boundary conditions. 

This mechanism of instability and stability of the solid-liquid interface and its movement to the upstream direction discussed above can be explained more quantitatively as follows.
The perturbed part of heat conservation equation (18) in Ref. \cite{4} is
\begin{equation}
L\frac{\partial \zeta}{\partial t}
=K_{s}\frac{\partial T'_{s}}{\partial y}\Big|_{y=\zeta}
-K_{l}\frac{\partial T'_{l}}{\partial y}\Big|_{y=\zeta}.
\label{eq:mech6}
\end{equation}
Taking imaginary part of Eq. (\ref{eq:mech6}), it becomes 
\begin{equation}
L{\rm Im}[\sigma \exp(\sigma t+ikx)]\zeta_{k}=q_{l}-q_{s},
\label{eq:mech7}
\end{equation}
where $q_{l}$ and $q_{s}$ are given by Eqs. (\ref{eq:mech3}) and (\ref{eq:mech4}).
The imaginary part of the left hand side of Eq. (\ref{eq:mech7}) can be written as
\begin{equation}
{\rm Im}[\sigma \exp(\sigma t+ikx)]
=|\sigma|\exp(\sigma_{r}t)\sin\left[k(x-v_{p}t)-\phi\right],
\label{eq:mech8}
\end{equation}
where $|\sigma|=\sqrt{\sigma_{r}^{2}+\sigma_{i}^{2}}$, and 
\begin{equation}
\sigma_{r}=|\sigma|\cos\phi,
\hspace{1cm}
\sigma_{i}=-|\sigma|\sin\phi.
\label{eq:mech9}
\end{equation}
Figure \ref{fig:hfluxk634,hfluxk953,hfluxk1200} (a)-(c) show that $\phi<0$. Noting that $v_{p}=-\sigma_{i}/k$, from the second equation of Eq. (\ref{eq:mech9}), when $\phi<0$, $\sigma_{i}$ is positive, therefore, $v_{p}<0$. From the first equation of Eq. (\ref{eq:mech9}), we find the sign of $\sigma_{r}$ and the corresponding figures as follows:
\begin{equation} 
\sigma_{r}\begin{cases}
                  >0 & \text{( $ -\frac{\pi}{2}< \phi < 0 $ )
                        Fig. \ref{fig:hfluxk634,hfluxk953,hfluxk1200}(a)} \\
                  =0 & \text{( $ \phi=-\frac{\pi}{2}$ ) 
                        Fig. \ref{fig:hfluxk634,hfluxk953,hfluxk1200}(b)} \\ 
                  <0 & \text{( $ -\pi < \phi < -\frac{\pi}{2}  $ )
                        Fig. \ref{fig:hfluxk634,hfluxk953,hfluxk1200}(c)}.
           \end{cases}  
\label{eq:mech10}
\end{equation}
Equation (\ref{eq:mech10}) indicates that unstable, neutral, and stable regions of the solid line in Fig. \ref{fig:growthrate} are completely consistent with Fig. \ref{fig:hfluxk634,hfluxk953,hfluxk1200} (a)-(c), and that the direction of phase velocity is the same as the solid line in Fig. \ref{fig:phasevelocity} if we restrict ourselves to the wavelength region observed on the surface of icicles or the inclined plane.

Likewise, if we apply Eq. (\ref{eq:dis8}) for $H_{l}(z)$ to Eqs. (\ref{eq:mech3}) and (\ref{eq:mech4}), $\phi>0$, therefore, we obtain $v_{p}>0$. This is consistent with the direction shown in the dashed and dotted lines in Fig. \ref{fig:phasevelocity}. According to the $O$-$F$ model, the stability of the solid-liquid interface is due to uniformalization of the temperature distribution along the layer by fluid flow \cite{5}. However, the fluid flow never make uniform the temperature distribution. If we give a correct interpretation for the stabilization of the solid-liquid interface in the $O$-$F$ model, which is essentially the same as that explained in our model. In the stable region, fluctuation of the solid-liquid interface and distribution of heat flux tend to be out of phase. At the protruded part, heat flux is small, while at the depression part, heat flux is large, therefore the flatness of the interface is restored. As a result, the dotted line in  Fig. \ref{fig:growthrate} is obtained. For the dashed line in  Fig. \ref{fig:growthrate}, out of phase between fluctuation of the solid-liquid interface and distribution of heat flux never occur. Therefore, this case is always unstable.

\section{Conclusion}
We have provided a physical interpretation for the morphological instability and stability of the solid-liquid interface occurring during a crystal growth from an undercooled thin parabolic shear flow of water on the surface of icicles or the inclined plane. The wavy pattern with a characteristic wavelength are observed on longer length scales compared to the one determined by the competition of the Mullins-Sekerka instability due to diffusion and stabilization due to the Gibbs-Thomson effect. We have found that phase difference between fluctuation of the solid-liquid interface and distribution of heat flux at the deformed solid-liquid interface, whose difference depends on the magnitude of restoring forces due to gravity and surface tension, is the cause for destabilization or stabilization of the interface, and that the direction of phase shift of the distribution of heat flux against the solid-liquid interface is related to the direction of migration of the solid-liquid interface.

\begin{acknowledgements}
The author would like to thank K. Iga for useful comments. The author would also like to thank Y. Ikoma for drawing some figures. 
\end{acknowledgements}

\end{document}